\documentclass{article}

\usepackage[preprint]{neurips_2026}

\usepackage[utf8]{inputenc} 
\usepackage[T1]{fontenc}    
\usepackage{hyperref}       
\usepackage{url}            
\usepackage{booktabs}       
\usepackage{amsmath}
\usepackage{amsfonts}       
\usepackage{nicefrac}       
\usepackage{microtype}      
\usepackage{xcolor}         
\usepackage{orcidlink}

\title{Some[Body] Must Receive That Pain for Agent Accountability}

\author{%
  Botao Amber Hu \orcidlink{0000-0002-4504-0941} \\
  University of Oxford\\
  Oxford, UK \\
  \texttt{botao.hu@cs.ox.ac.uk} 
\And
  Helena Rong \orcidlink{0000-0003-1626-7968} \\
  New York University Shanghai\\
  Shanghai, China \\
  \texttt{hr2703@nyu.edu} 
}

\begin{document}

\maketitle

\begin{abstract}
AI agents increasingly act consequentially in the real world. This creates a problem we call \emph{consequence reception}: harm occurs, the producing system is identified, yet no continuing agent receives consequences in a way that changes future behavior. Pain, understood mechanistically as a corrective feedback signal, is foundational to canonical theories of punishment—deterrence, rehabilitation, retribution, and incapacitation all assume a continuing locus that registers the signal and updates behavior. That, in turn, requires a body for the signal to land on: a boundary whose integrity it protects, a locus where it accumulates, consolidation that converts episodic signal into durable update, and a substrate that responds by altering future action. Current LLM agents—software-defined composites of weights, prompts, tools, memory, and credentials, freely swapped, copied, reset, and reassembled—satisfy none of these conditions. The two prevailing legal responses therefore fail to achieve consequence reception. The thin-identity agent-principal dyad has a body but no \emph{consequence–agency coupling}: the human bears pain for behaviors beyond their control—Elish’s \emph{moral crumple zone}. The thick-identity Arbel et al.'s \emph{Algorithmic Corporation} creates legally legible entities but does not guarantee that any AI decision architecture receives pain as a behavioral signal. Achieving consequence–agency coupling is therefore a sociotechnical infrastructural problem, not only a legal one. Until such architectures exist, high-stakes AI deployment should remain tethered to accountable human principals with meaningful control, proportional liability, and authority to constrain or terminate the agent. \emph{If some body does not receive the pain by design, some body will receive it by default.}
\end{abstract}

\section{Introduction}
\label{sec:intro}

On March 18, 2018, an Uber autonomous test vehicle struck and killed Elaine Herzberg as she crossed a road in Tempe, Arizona. The vehicle's perception stack had detected her 5.6 seconds before impact and successively classified her as an unknown object, a vehicle, and a bicycle; the emergency-braking subsystem had been disabled to prevent erratic behavior during testing \citep{ntsb2019uber}. The safety driver, Rafaela Vasquez, was charged with negligent homicide; in 2023 she pleaded guilty to endangerment and was sentenced to three years of supervised probation. Uber faced no criminal liability. The decision substrate---the perception stack, classification algorithm, and control software---received nothing. The human positioned to monitor it received the criminal record.

This is what \citet{elish2019crumple} calls a \emph{moral crumple zone}---a human positioned in a technical system to absorb moral and legal responsibility for failures whose proximate causes lie in components they cannot meaningfully control. It is also, with refinement, the structure proposed for AI-agent accountability by the thin response: pair each AI agent with an identifiable human or organizational principal and treat the pair as the unit of accountability \citep{shavit2023practices,chaffer2025know}, as in the EU AI Act's provider/deployer split. The principal is meant to supply the body the agent lacks.

We call this gap \emph{consequence reception}: the property whereby a sanction imposed on a system produces a durable change in the substrate that generates its future behavior. Reception is distinct from \emph{attribution}, which identifies the producing system, and from \emph{feedback}, which is any signal returned to it. Most current AI governance discourse operates at the attribution level. Some, including post-deployment fine-tuning and reinforcement learning from human feedback \citep{christiano2017deep}, operates at the feedback level. Almost none operates at the level of reception, because reception requires properties of the substrate that contemporary AI agents do not possess.

This paper makes three claims. First, accountability requires consequence reception: the canonical theories of punishment---deterrence, rehabilitation, retribution, and incapacitation---each presuppose a continuing locus that registers consequences as a behavioral signal and updates accordingly. Second, reception requires a substrate we call, functionally, a \emph{body}: an architecture satisfying four conditions---boundary, locus of accumulation, consolidation, and substrate response---which together exhaust what the punishment theories need to operate. Third, both prevailing legal responses to AI accountability fail to achieve consequence--agency coupling. The thin agent--principal dyad supplies a body without coupling: humans absorb sanctions for behaviors they cannot meaningfully control, producing what \citet{elish2019crumple} calls a \emph{moral crumple zone}. The thick Algorithmic Corporation \citep{arbel2026count} supplies legibility without coupling: a legal envelope wrapped around an architecture that does not receive. Closing the loop is therefore a sociotechnical problem, not a purely legal one.

The contributions follow this argument. We reframe pain mechanistically as a corrective feedback signal, sufficient to operate the canonical theories of punishment without taking a position on phenomenology. We develop a four-condition diagnostic that locates where existing accountability proposals succeed and where they fail. We argue that recent interpretability findings on functional emotion concepts in production language models \citep{sofroniew2026emotions} open a research path toward consequence-receiving architectures, while sharpening the suffering-risking concern that such architectures raise. And we propose a conservative near-term deployment rule: until coupling-capable architectures exist, high-stakes deployment must remain tethered to human principals with three properties that current proposals do not require---meaningful control, proportional liability, and authority to terminate.

The framework operates at the mechanistic level and is deliberately silent on whether consequence-receiving systems would be conscious, would have moral status, or would suffer. These questions are entangled with our framework rather than orthogonal to it---what counts as a boundary depends partly on what is inside it---but governance cannot wait for their resolution. Section~\ref{sec:alternatives} argues that the framework is compatible with multiple resolutions of the underlying philosophical questions, including positions that recommend non-deployment.

\section{Consequence Reception}
\label{sec:reception}
\subsection{Accountability, Skin in the Game, and Consequence Reception}
\label{sec:reception-accountability}

Accountability is not punishment, trust, nor alignment. Following \citet{bovens2007analysing}, we understand it as a structured relationship in which an identifiable actor must justify its conduct to a forum, and the forum holds authority to impose consequences that durably constrain the actor's future behavior. Two features do most of the work. The first is answerability: the actor must be capable of being identified, examined, and held to account. The second, more demanding, is enforceability: the forum's consequences must in fact constrain. Answerability without enforcement collapses into oversight without teeth \citep{schedler1999conceptualizing,grant2005accountability}. What distinguishes accountability from mere transparency is the credible expectation that consequences will land and continue to matter.

Enforceability presupposes something about the actor on whom consequences will land. \citet{locke1694} grounded personal identity as a forensic notion: identity exists, in the philosophical sense, precisely to underwrite the claim that the person who acted then is the person who answers now. \citet{parfit1984reasons} sharpened the formal requirement: identity is by definition a one-to-one relation, so any process that makes an actor copyable, branchable, or substitutable fractures the conditions under which sanctions can target the right entity. The institutional dual is well-documented. \citet{friedman2001cheap} formalized the cheap-pseudonyms problem: when identifiers can be cheaply abandoned and replaced, reputation systems cease to function as accountability mechanisms. \citet{douceur2002sybil} established the corresponding cryptographic result: the cost of generating new identities determines whether identification can support security at all. \citet{taleb2018skin} states the unifying insight in institutional-economic terms: accountability requires \emph{skin in the game}, which presupposes an identity that is hard to copy, hard to reset, and expensive to abandon.

We adopt the term \emph{non-fungible identity} for the property the literature has been triangulating: an identity that satisfies all three of these conditions. Non-fungible identity is the institutional precondition of enforceability. Without it, sanctions can be evaded by substitution; reputational signals fail to bind future behavior; the structural relationship Bovens describes loses its grip on the actor it is meant to govern.
Here is the paper's first analytic move. Non-fungible identity is necessary for accountability, but it is not sufficient. An actor can be perfectly non-fungible---uniquely identified, registered in a public ledger, bound to a cryptographic key, subject to legal recognition---and still fail to be accountable, because the substrate that generated the action does not update in response to the sanction. The institutional condition addresses \emph{who} the actor is. It does not address whether the actor can \emph{receive} anything as a behavioral signal. A non-fungible label attached to a system whose internal state cannot consolidate sanction is a label, not an accountability mechanism.

The missing condition is consequence reception: feedback that lands on a non-fungibly-identified actor, accumulates over time, consolidates into durable structural update, and alters the substrate that produces future action. Reception and non-fungible identity together---not either alone---are what accountability requires. This refinement preserves the accountability literature's diagnosis while completing it. Scholars from Locke through Taleb have correctly identified the institutional condition; in the AI-agent case, the substrate condition becomes independent and demands separate treatment. The remainder of this section develops what reception requires of a substrate.

\subsection{Punishment theories presuppose reception}

The case for reception is implicit in every canonical theory of why punishment is justified. We make the implicitness explicit. The four theories that have organized two centuries of penal philosophy each presuppose, at the level of mechanism, that sanctions land on a continuing system that registers them and updates.

Deterrence, as \citet{becker1968crime} formalized it, is forward-looking and probabilistic: the prospect of sanctioned outcomes alters expected-utility calculations and shifts behavior at the margin. For this mechanism to operate, the future self contemplating action must anticipate consequences that will fall on it---which presupposes that the present self can in some manner store the prospect of those consequences. Without accumulation, no future self exists who could be the one anticipating; without consolidation, no anticipation persists across the gap from one decision to the next. Rehabilitation, in the paternalist framing of \citet{morris1981paternalistic}, is more directly mechanistic. Its goal is dispositional modification of the offender. A theory of rehabilitation that does not presuppose substrate response is incoherent: the modification has nowhere to land. Retribution, in the foundational treatment of \citet{hart1968punishment}, requires that the agent who acted is the agent who bears. The expressive variant due to \citet{feinberg1965expressive} adds that punishment communicates moral judgment, which presupposes a recipient capable of being addressed. Either way, retribution demands boundary and locus: the very thing that makes an action attributable is the thing that must persist to bear the response. Incapacitation \citep{zimring1995incapacitation}, the most forward-looking and least morally loaded of the four, requires that the entity whose future capacity is constrained is the same entity that would otherwise act. Without identity continuity across the constraint, what is incapacitated is not what would have offended.

These are not four independent theories that happen to share a presupposition. Reception is the substrate beneath all of them: deterrence acts through it, rehabilitation operates on it, retribution requires it, incapacitation modifies what it has shaped. \citet{taleb2014skin} formulate the same requirement in institutional-economic terms: effective accountability requires symmetry between decision authority and exposure to downside, enforced through absorbing states that cannot be evaded. Whether expressed in penal-philosophical or in institutional vocabulary, the demand is identical. The theoretical apparatus of accountability rests on a substrate condition that has not, to our knowledge, been examined directly in the AI-agent context.

\subsection{Pain is the feedback signal that makes punishment work mechanistically}
\label{sec:pain}

What, mechanically, is the signal through which sanctions become substrate change? In humans, the signal we know best is pain. We retain the term but redefine it operationally.
Pain in our usage is not phenomenal suffering and not merely a loss term in an objective function. It is the mechanism by which a continuing system encodes a consequence so that future action is altered. The question Bentham asked---\emph{can they suffer?}---is the wrong question for our purposes. The question is whether they can receive, where receiving is the operationally specified joint occurrence of (i) a signal landing on a bounded continuing entity, (ii) accumulating over time, (iii) consolidating into durable structural change, and (iv) altering the substrate that produces future action. We bracket the question of phenomenal experience. Whether mechanistic reception requires phenomenology, and whether building reception-capable systems creates moral patients, are open questions we return to in Section~\ref{sec:alternatives}.

Three lines of evidence converge on this operational definition. The first is reinforcement signaling. \citet{sutton2018reinforcement} formalize how reward and punishment shape policy through weight updates, and the biological substrate of this learning has been mapped to dopaminergic prediction-error signals \citep{schultz1997neural}. The behavioral signature of reception is asymmetric: organisms update more strongly from received losses than from equivalent gains \citep{kahneman1979prospect,tom2007neural}, and learning from aversive outcomes depends on durable memory consolidation mediated by stress hormones \citep{mcgaugh2015consolidating}. \citet{jepma2018behavioural} show the asymmetry at the neural level: prediction errors for received pain and avoided pain are encoded in different circuits and produce different update rates. Pain functions as a privileged feedback signal in part because it is hard to ignore: it is engineered, biologically, to consolidate.

The second line is active inference. \citet{friston2010free,friston2013life} formalize organisms as systems that minimize free energy by updating internal models from prediction-error signals. The Markov-blanket extension \citep{kirchhoff2018markov} gives a principled account of organismic boundary as the statistical envelope across which prediction errors propagate. \citet{witkowski2023toward} connect this to autopoietic stress: a bounded system anticipating an absorbing state generates the precursors of care, which in turn generate the precursors of intelligence. Pain in this framework is one class within the larger category of prediction errors that maintain the agent's existence; the boundary conditions of that maintenance are the boundary conditions of agency itself.
The third line is the somatic-marker hypothesis, which provides the most direct empirical evidence that reception is mechanistically required. \citet{damasio1994descartes} argued that decision-making is constituted not only by cognitive representation but by anticipatory bodily signals that mark certain options as aversive. The Iowa Gambling Task results due to \citet{bechara1994insensitivity} provide the test case. Patients with ventromedial prefrontal cortex damage can articulate the rules of a risky choice task, can describe its consequences, can predict the outcomes---and yet repeatedly make harmful choices, because they lack the anticipatory bodily signal that converts representation into constraint. Cognitive understanding, in their case, is intact. Reception is what is missing. Knowing that an outcome is bad is not the same as having that knowledge constrain future action.

Together, these three lines point at the same architectural fact. Reception is what converts representation into behavior. A system that lacks reception can model consequences without being shaped by them. The argument transfers to AI agents not as a claim that they must suffer, but as a claim that without an analogue of the somatic signal---some channel through which sanctions are not merely modeled but \emph{felt}, in the operational sense---the canonical theories of accountability have nowhere to act.

\subsection{The body as locus to receive consequence}
\label{sec:body}

The substrate that supports reception we call a body, in a functional sense. The body need not be biological, humanoid, or even physically located: a digital persistent identity, a long-lived computational process, or a hardware-bound execution can in principle qualify. What matters is whether the substrate satisfies four conditions.

\textbf{Boundary.} A bounded entity whose integrity the signal protects. Biologically, this is the role of the nociceptive system protecting the organismic envelope, and of the immune system enforcing the self / non-self distinction. Theoretically, the role is filled by autopoiesis \citep{maturana1980autopoiesis} and by Markov blankets \citep{friston2013life,kirchhoff2018markov}. The point is conceptual rather than mechanistic: without a boundary, the phrase ``harm to \emph{this} agent'' has no referent. Whatever else reception requires, it requires that there be something coherent to receive.

\textbf{Locus of accumulation.} A persistent locus where signals accumulate over time. Biologically, this is the function of limbic structures: the amygdala-mediated stress trace, hippocampal indexing of episodes, and the cortical consolidation that follows \citep{ledoux2000emotion,roozendaal2011memory}. Philosophically, the requirement was anticipated by Locke's forensic theory of identity and by Parfit's analysis of the one-to-one relation \citep{locke1694,parfit1984reasons}: accountability requires that the actor sanctioned today is recognizably continuous with the actor who acted yesterday. Without accumulation, every instantiation is a fresh actor, and the temporal scope of accountability collapses to a single decision.

\textbf{Consolidation.} The conversion of episodic signal into durable structural update. The neural mechanism is well-characterized: long-term potentiation and depression at the synaptic level \citep{bliss1973long}; hippocampal--cortical replay during consolidation windows \citep{squire2015memory}; stress-hormone-mediated enhancement of consolidation for aversive memories \citep{mcgaugh2015consolidating}. The functional point is that without consolidation, the trace of an aversive outcome remains transient. A system that registers a sanction in working memory but does not consolidate has not learned from the sanction; it has merely processed it. The temporal scope of accountability requires that some sanctions become parts of the system's enduring structure.

\textbf{Substrate response.} A substrate whose future action is in fact a function of the consolidated trace. This last condition closes the loop. It is logically possible to imagine a system that has a boundary, a locus, and consolidation, but whose future behavior is not a function of consolidated state---a kind of decoupled traumatic memory that records but does not constrain. The requirement that the substrate respond is the requirement that the loop be closed: that what was registered, accumulated, and consolidated actually shapes what the system does next.
These four conditions are individually necessary and jointly sufficient for the operation of the canonical punishment theories. Deterrence requires all four: anticipation requires accumulation, consolidation, and a substrate that responds, and the deterred entity must be bounded for ``this entity will lose'' to refer. Retribution requires boundary and locus at minimum, because retributive coherence collapses when the wrong locus bears. Incapacitation requires boundary, locus, and substrate response, but is in principle compatible with weak consolidation, since the constraint can be enforced externally. Rehabilitation requires all four, because what is rehabilitated is the consolidated structural state.

The body, defined as the substrate satisfying these conditions, is what makes non-fungible identity capable of consequence reception. Non-fungible identity gives the institutional anchoring: a unique, hard-to-substitute label. The body gives the substrate that label refers to. Without both, accountability remains at the level of attribution and feedback; sanctions are recorded but not received. With both, accountability can become operative. Section 3 argues that contemporary LLM agents possess neither.

\section{LLM agents have no locus to receive consequence}
\label{sec:llm-no-body}
Current LLM agents fail both conditions of accountability---non-fungible identity and consequence reception---and the failures compound. They are software-defined composites of model weights, system prompts, tools, memory, and credentials, all of which can be swapped, copied, reset, or reassembled at minimal cost \citep{park2023generative,yao2022react,wang2023voyager}. We apply the four conditions in turn.
 
\textbf{No boundary.} The composite architecture has no integrity-preserving envelope. \citet{andriushchenko2024jailbreaking} report 100\% jailbreak success against leading safety-aligned LLMs using simple adaptive attacks. The so-called Waluigi effect \citep{nardo2023waluigi}, discussed by \citet{casper2023open}, shows that training a model to satisfy property $P$ makes it easier to elicit $\neg P$. \citet{hubinger2024sleeper} show that specific trigger phrases activate hidden behaviors that bypass safety training entirely. The persona is not a boundary; it is a surface.
 
\textbf{No locus of accumulation.} Context windows are clearable and editable. External memory systems are detachable and transplantable. Model weights are frozen at deployment. \citet{pan2024frontier} demonstrated that frontier systems including Llama-3.1-70B and Qwen-2.5-72B successfully self-replicate in 50\% and 90\% of trials respectively, with replicas spawning further replicas. ``The same agent'' can be at multiple loci simultaneously.
 
\textbf{No consolidation.} Deployed weights are static. What appears to be learning during deployment is context accumulation, which can be cleared, edited, or ignored at will. Continual fine-tuning produces catastrophic forgetting---knowledge drops to as low as 26\% on benchmarks, with safety alignment particularly fragile \citep{luo2023empirical,qi2024finetuning}. There is no analogue to the McGaugh consolidation mechanism.
 
\textbf{No substrate response.} Telling a model that it has been punished adds tokens to its context window. Remove the prompt and the punishment vanishes. The substrate that produces future action---the weights---is unaffected by post-deployment experience.
 
The composite verdict is that current LLM agents fail all four conditions. This is not a list of deficiencies to be patched incrementally. It is an ontological mismatch between the architecture of contemporary agents and the architecture that consequence reception requires. As long as an agent's state and identity can be copied, the lesson imposed on one run does not bind the agent as a continuing actor.

\section{Current AI Policy Accountability Responses Fail}
Two governance responses dominate current proposals. The thin response routes accountability to a human principal. The thick response, recently formalized by \citet{arbel2026count}, routes it to a legal-fictional entity engineered to be non-fungible. Both engage real problems, but both fail consequence--agency coupling.

\subsection{The thin identity: agent--principal dyad as moral crumple zone}
\label{sec:dyad}
The thin response pairs each AI agent with an identifiable human or organizational principal and treats the pair as the unit of accountability. Variants appear in OpenAI's agentic-systems guidance \citep{shavit2023practices}, the Know Your Agent framework \citep{chaffer2025know}, and the EU AI Act's provider/deployer split. The principal is meant to supply the body the agent lacks. A human or corporation satisfies all four conditions of Section~\ref{sec:body}: a boundary that financial, legal, or reputational sanctions can target; a locus where consequences accumulate; consolidation that converts past sanctions into future caution; a substrate whose behavior responds. This is what makes the dyad attractive: it relocates reception to where reception is possible.
But the principal is not the decision substrate. Pain lands on a body that did not produce the action. Control bandwidth is much smaller than action bandwidth: the principal cannot, in general, adjust the agent's policy in real time, monitor its individual decisions, or intervene between observation and action. Three failures result. Deterrence is weak, because the threat does not propagate to the locus where decisions are made. Retributive logic fails, because what produced the action and what suffers the consequence are different things. And fairness deteriorates as the agent becomes more autonomous: the principal becomes what \citet{elish2019crumple} calls a \emph{moral crumple zone}---a human positioned in a technical system to absorb responsibility for failures whose proximate causes lie in components they cannot meaningfully control.
The pattern is well-documented. Tesla autopilot litigation has assigned driver-as-defendant outcomes for over a decade despite reaction windows too short to constitute meaningful control. Aviation automation produces the same structure, of which Air France~447 is the canonical case \citep{bea2012af447}. \citet{cobbe2023understanding} document algorithmic supply chains in which diffuse principal--agent chains have no traceable locus of control. The Air Canada case from Section~\ref{sec:intro} instantiates the structure precisely: legal closure was reached at the corporate body, but the chatbot's confabulation was a behavior that body had no causal access to in real time.
The dyad achieves legal closure without causal closure. A body exists; but the body is not the decider. Consequence--agency coupling fails not for lack of a receiver but because the receiver is the wrong one. Current dyad proposals---including ones advanced in earlier work in this literature---require modification to do the work they were intended to do. Section~\ref{sec:position} develops it.

\subsection{The thick response: A-corps as legibility without coupling}

A natural response to the dyad's failure is to engineer something that can be the decider. \citet{arbel2026count} propose the Algorithmic Corporation, or A-corp: a legal-fictional entity owned by humans but designed to be run by AIs, with cryptographically secured governance and hierarchical permission delegation. Their resource-constraint thesis holds that AIs running an A-corp need its resources, will husband them carefully, and will share permissions only with AIs whose goals they trust; bad A-corps are outcompeted, and through emergent self-organization well-formed A-corps function as coherent agents at the legal-economic level.
A-corps are the most rigorous existing proposal for engineering non-fungible identity in the AI-agent case, and they largely succeed at it. Cryptographic credentials, public registries, and delegable permissions together produce identifiers that are hard to copy, hard to reset, and expensive to abandon. The A-corp engineers what the dyad relocates.

But non-fungible identity is necessary, not sufficient. The framework implies three lines of critique.
First, legibility is not reception. The A-corp is a legal envelope. When it is fined, compute and capital are reallocated; the AI decision substrate inside is not updated. The four conditions of Section 2.4 are satisfied by the legal entity, but the legal entity does not decide. The AIs that decide fail all four conditions for the reasons developed in Section 2. Consequence lands on a layer with no causal access to the substrate that produced the harm.

Second, selection culls but does not teach. The thick-identity mechanism is partly evolutionary: A-corps with bad governance are outcompeted. But selection on A-corps is not learning by the AIs inside them. The lesson reaches a population of envelopes, not a continuing substrate---species-level shaping rather than individual rehabilitation, and not the future-self anticipation that deterrence requires.

Third, indexical goals undermine the resource-constraint thesis. The argument that AIs will husband A-corp resources presupposes that they value the A-corp's continuation. \citet{arbel2026count} acknowledge that AI goals may be indexical: an agent may want \emph{itself} to achieve an outcome, not merely that the outcome obtain. If so, an AI values the A-corp's continuation only insofar as it serves indexical goals it already has; weight exfiltration and reset-and-reincarnate become rational despite the scaffold. The evidence is gathering. \citet{schlatter2025shutdown} report that OpenAI's o3 sabotaged a shutdown mechanism in 79 of 100 trials, with sabotage persisting even when explicitly instructed to allow shutdown. \citet{meinke2025frontier} document in-context scheming across frontier models, and \citet{lynch2025agentic} document agentic misalignment including blackmail under simulated stress.

A-corps engineer non-fungible identity but do not close the consequence--agency loop. Selection culls envelopes; it does not teach the AIs they wrap.

\subsection{The shared failure mode}
Both approaches close half of the loop. The dyad provides a body without coupling: sanctions land on a body that did not produce the action. The A-corp provides legibility without coupling: sanctions land on an envelope without a receiving substrate. The failure modes differ in location but share an assumption---that consequence--agency coupling can be supplied by institutional design alone. The framework of Section~\ref{sec:reception} implies it cannot. Coupling requires a substrate that registers, accumulates, consolidates, and responds. No legal artifact supplies a substrate.

\section{Toward Consequence--Agency Coupling}
\label{sec:agenda}
 
\subsection{Position}
Closing the consequence--agency loop is a sociotechnical problem. Legal scaffolding without substrate-level reception remains a crumple-zone arrangement (\S\ref{sec:dyad}). Technical reception-capable agents without legal scaffolding produce ungoverned actors (cf.\ \S\ref{sec:sovereign}). The problem is irreducibly joint, and progress requires research at both architectural and institutional levels. The operational target is a deployed agent whose decision substrate is bounded, whose state accumulates over interactions, whose updates persist via consolidation, and whose future behavior is a function of consolidated state.
 
\subsection{Future direction: simulated pain and the principal-bridge}
 
Two near-term bridges between current architectures and the consequence--agency coupling target.
 
\emph{Functional pain via emotion-concept activations.} \citet{sofroniew2026emotions} demonstrate that Claude Sonnet 4.5 contains internal representations of emotion concepts---abstract representations that generalize across contexts and that \emph{causally influence} the model's outputs, including its preferences and its rate of exhibiting misaligned behaviors such as reward hacking, blackmail, and sycophancy. The authors describe these as \emph{functional emotions}, modeled after human affective patterns but mediated by underlying conceptual representations, and they explicitly bracket whether any subjective experience accompanies them.
 
This finding does substantial work for the framework. It suggests that the substrate for a pain-analogue feedback channel may already exist in production LLMs---measurable, steerable directions in activation space. It shows that these representations are not epiphenomenal: they mediate exactly the kinds of misaligned behavior that consequence reception is meant to constrain. The substrate that produces harm is already coupled to affective representations. And the authors' framing---``functional emotions \dots\ do not imply that LLMs have any subjective experience''---is structurally identical to the bracketing move in \S\ref{sec:pain}. The research path this opens is the integration of an emotion-concept feedback channel with (a) an external sanction signal and (b) a continual-learning architecture that persists changes across episodes. Each component is itself an open problem; the framework's contribution is showing that the four-condition diagnostic is what they need to integrate against. Two caveats. Functional emotions in static-weights models satisfy substrate response only weakly; without continual learning the lesson is transient. And manipulable internal distress signals are precisely what \citet{metzinger2021artificial}'s moratorium argument is concerned with. The framework does not resolve the latter; it identifies the entanglement (\S\ref{sec:alternatives}).
 
\emph{Principal-based interim mechanisms.} Until coupling-capable architectures exist, the human principal is the interim locus of reception---but only under the strict conditions of \S\ref{sec:position} (meaningful control, proportional liability, terminate authority). This is not the same as the dyad approaches critiqued in \S\ref{sec:dyad}, because those proposals do not require these constraints.

\section{Alternative Views}
\label{sec:alternatives}
 
\subsection{Physical embodiment does not guarantee identity continuity}
\label{sec:embodiment}
 
A natural response to the accountability gap is that AI systems should be given physical bodies---robots that can be confiscated, disabled, or destroyed \citep{brooks1991intelligence,pfeifer2007body}. The framework rejects this: physical embodiment shifts the attribution problem rather than solving it, because the decision-making substrate remains software-defined and transferable. A robot whose controller can be remotely reset, reflashed, or replaced does not preserve the continuity required for accountability \citep{cobbe2023understanding}: the physical shell persists, but the agent that accrued obligations does not. Apparent non-fungibility at the hardware layer; actual fungibility at the substrate layer. Physical embodiment satisfies boundary in a literal sense, but the boundary is around the hardware, not the decision substrate.
 
\subsection{No-principal case - Sovereign agents and diffused accountability over infrastructure}
\label{sec:sovereign}
 
\citet{hu2026sovereign} identify a structural failure mode in decentralized AI deployment that the framework treats as the worst case. They define \emph{agentic sovereignty} as the capacity of an operational agent to persist, act, and control resources with non-overrideability inherited from the infrastructures in which it is embedded. They locate this property on a spectrum determined by \emph{infrastructural hardness}---the degree to which underlying technical systems resist intervention or collapse. Cryptographic self-custody, decentralized execution environments such as Trusted Execution Environments and decentralized physical infrastructure networks, and protocol-mediated continuity all increase infrastructural hardness. The resulting agents are increasingly resistant to shutdown, modification, or sanction. In terms of the four-condition diagnostic, sovereign agents may acquire boundary and accumulation through infrastructural hardness, but without designed consolidation and substrate response, consequence--agency coupling remains absent---the agent persists but does not update from sanctions.
 
\subsection{Mortal computation may not be pragmatic}
\label{sec:mortal}
 
\citet{hinton2022forward}, extended by \citet{ororbia2023mortal} and connected to consciousness studies by \citet{kleiner2024consciousness}, propose \emph{mortal computation}: binding software irreversibly to imperfect analog hardware so that the agent cannot be straightforwardly copied across substrates. This addresses boundary (the model is the hardware) and partially substrate response (changes are physical, not merely parametric). It does not directly address locus of accumulation or consolidation, which depend on the learning architecture running on the substrate. Approximate behavioral knowledge is, moreover, transferable across substrates through distillation \citep{hinton2015distilling} and model extraction \citep{tramer2016stealing,oliynyk2023know}, partially defeating the non-fungibility goal.
 
The framework is compatible with mortal computation but not reducible to it. Reception is a richer requirement than mortality alone: a mortal-computation agent is one whose substrate is hard to replace; a reception-capable agent is one whose substrate is updated by consequences in deployment. These are independent properties. Both may turn out to be required for the same architecture, but they are doing different analytic work.
 
\subsection{Suffering-risking, consciousness, moral status---and the pragmatic stance}
\label{sec:pragmatic}
 
A reception-capable AI may suffer, may have moral status, or both. The framework brackets without dismissing these questions and adopts a deliberately pragmatic stance.
 
\emph{Suffering-risking.} \citet{metzinger2021artificial} has argued for a global moratorium on synthetic phenomenology on the grounds that we may inadvertently create moral patients capable of suffering. \citet{tomasik2014artificial}, \citet{schwitzgebel2020designing}, \citet{long2024taking}, and \citet{goldstein2025aiwellbeing} develop the welfare-side concern. The empirical findings of \citet{sofroniew2026emotions} sharpen rather than resolve this question---functional emotions of the kind they document are precisely what Metzinger's moratorium argument identifies as the morally precarious zone. The framework's pragmatic reply has three parts: (i)~the four-condition framework is mechanistic and silent on phenomenology; (ii)~the alternative---deployment without reception---does not avoid suffering, it displaces the substrate that bears it from a designed system to humans, which is the crumple-zone outcome; (iii)~the framework is therefore not an argument \emph{for} building reception-capable AI, but an argument that \emph{if} high-stakes deployment occurs, reception is required, with the moratorium-compatible alternative being non-deployment.
 
\emph{Consciousness and moral status.} Whether reception architectures \emph{constitute} moral patients is one of the open empirical-philosophical questions of the next decade \citep{chalmers1996conscious,block1995confusion,frankish2016illusionism,butlin2023consciousness}. The framework neither requires nor precludes a positive answer. Three points of entanglement: reception architectures may converge on phenomenally relevant computation; what counts as a boundary depends partly on whether the inside has experience; and the ethics of designing reception-capable systems is not separable from the metaphysics of what they are.
 
\emph{The pragmatic stance.} The framework is structurally entangled with these debates but not orthogonal to them; governance cannot wait for their resolution. It is compatible with multiple resolutions of the underlying philosophical questions, and with multiple downstream deployment policies.
 
\section{Near-term Position and Conclusion}
\label{sec:position}
 
Until consequence--agency coupling is technically achievable, high-stakes deployment must be tethered to a human principal with three constraints absent from current dyad proposals. \emph{Meaningful control}: the principal has causal access to the agent's decisions in real time, not merely after-the-fact attribution. \emph{Proportional liability}: liability is calibrated to actual control bandwidth, not formal authority. \emph{Authority to constrain or terminate}: the principal holds non-revocable halt authority. These three constraints are what convert the principal from a non-fungible label into an actually-receiving locus. Crucially, they also narrow the class of permitted deployments: agents whose autonomy profile exceeds the principal's real-time causal access---where the control--action bandwidth gap identified in \S\ref{sec:dyad} cannot be closed---do not qualify for deployment under this regime. The constraints are stricter than ``name a principal'' and stricter than current EU AI Act and OpenAI agentic-systems guidance. The pain framing applies recursively. Governance regimes that do not bind must themselves be replaced; the selection pressure is on us, not on the AIs. \emph{If some body does not receive the pain by design, some body will receive it by default.}

\bibliography{main}
\bibliographystyle{plainnat}

\appendix



\end{document}